\documentclass[prd,preprintnumbers,12pt]{revtex4}
\pdfoutput=1
\usepackage{epsfig}
\usepackage{graphicx}
\usepackage{bm}
\usepackage{amsmath}
\usepackage{amsfonts}
\usepackage{amssymb}

\newcommand{\be}{\begin{equation}}

\newcommand{\ee}{\end{equation}}
\newcommand{\bea}{\begin{eqnarray}}
\newcommand{\eea}{\end{eqnarray}}

\newcommand{\nn}{\nonumber}
\newcommand{\de}{\partial}
\def\nn{\nonumber}
\def\de{\partial}

 \def\slash#1{\setbox0=\hbox{$#1$}#1\hskip-\wd0\dimen0=5pt\advance
       \dimen0 by-\ht0\advance\dimen0 by\dp0\lower0.5\dimen0\hbox
         to\wd0{\hss\sl/\/\hss}}
\def\be{\begin{equation}}

\def\ee{\end{equation}}

\def\bea{\begin{eqnarray}}
\def\eea{\end{eqnarray}}

\def\7{\tilde}
\def\8{\hat}

 \def\slash#1{\setbox0=\hbox{$#1$}#1\hskip-\wd0\dimen0=5pt\advance
       \dimen0 by-\ht0\advance\dimen0 by\dp0\lower0.5\dimen0\hbox
         to\wd0{\hss\sl/\/\hss}}

\usepackage{color}
\usepackage{color}

\newcommand{\eq}[1]{(\ref{eq:#1})}
\def\seta{\eta^*}
\def\LL{{\cal L}}

\begin{document}

\title{
Translational symmetries of quadratic lagrangians }

\author{Andrea Barducci}\email{barducci@fi.infn.it} \author{Roberto Casalbuoni} \email{casalbuoni@fi.infn.it} \affiliation{Department of Physics and Astronomy, University of Florence and INFN, Via G. Sansone 1, 50019 Sesto Fiorentino (FI), Italy}

\begin{abstract}
 In this paper we show that a quadratic lagrangian,  with no constraints, containing  ordinary time derivatives up to the order $m$ of $N$ dynamical variables, has $2mN$ symmetries consisting in the translation of the variables with solutions of the equations of motion. We construct explicitly  the generators of   these transformations  and prove that they satisfy   the Heisenberg algebra. We  also analyse other  specific cases which are not included in  our previous statement: the Klein-Gordon lagrangian,  $N$ Fermi oscillators and the Dirac lagrangian. In the first case, the system is described by an equation involving partial derivatives,  the second case is described by Grassmann variables and the third  shows both features.  Furthermore, the Fermi oscillator and the Dirac field  lagrangians lead to second class constraints. We prove that also in these last two cases there are translational symmetries and we construct the algebra of the generators. For the Klein-Gordon case we find a continuum version of the Heisenberg algebra, whereas in the other cases, the  Grassmann generators satisfy, after quantization, the algebra of the Fermi creation and  annihilation operators.

 \end{abstract}
\maketitle

\section{Introduction}

A fundamental feature of a system of differential linear equations is that any linear combination of solutions is a solution. This implies that translating the dynamical variables with solutions of the equations leaves the equations invariant. If the system is of degree $d$ and  is composed of  $N$ independent 
  equations, the number of  translations leaving invariant the system is equal to $dN$. In the case in which the system can be described by a lagrangian (which is necessarily quadratic), two questions arise. The first one is: are the  previous invariances of the equations of motion  also invariances of the lagrangian? And the second question:   if the lagrangian is  invariant,  which are the generators of these transformations and which is their algebra? In this paper we will answer  these two questions showing that the translation of the variables with a solution of the equations of motion is a symmetry of the lagrangian up to a total derivative. Furthermore we will prove that, if there are no constraints, there are $dN$ symmetries and will explicitly construct  the generators. We will also show  that their Poisson algebra is the same as that of  the canonical variables, implying that after quantization they satisfy the Heisenberg algebra.

 There are various special cases of these symmetries. The most obvious case is the invariance under spatial translations and Galilei boosts of the non relativistic free particle. In fact, both these transformations can be seen as translations of the coordinates with solutions of the equations of motion of the free particle  (see later). The case of the harmonic oscillator has been studied in \cite{Lutzky:1978vi}. Also the case of the Pais-Uhlenbeck \cite{Pais:1950za} oscillator has been considered in \cite{Andrzejewski:2014rza} and \cite{Masterov:2015ija}. In this case the lagrangian depends on higher order derivatives. In these last two cases the translational symmetries are a subgroup of the Newton-Hooke symmetries that were introduced in \cite{Bacry:1968zf}
 where the translational symmetries were called inertial transformations. The Newton-Hooke symmetries have also been applied  to cosmological problems  in \cite{Derome:1972zf,Dubois:1973dj}.
 
 We will show that in the case of a lagrangian depending on $m$ time derivatives of the $N$ dynamical variables, the number of conserved generators is equal to that of independent solutions of the equations of motion, namely $2mN$, where $2m$ is the order of the equations of motion,  and that the corresponding symmetry algebra is universal independent of the details of the lagrangian. 
 Only the transformations of the dynamical variables depend of the details of the lagrangian, since the parameters of the translations are the solutions of the equations of motion. This universal algebra  turns out to be  the Heisenberg algebra.
 
 Although we have done a general analysis only for a system of $N$ degrees of freedom without constraints and satisfying an ordinary differential equation of order $2m$, we have also considered  three particular cases which show that, most probably, our results are generally  valid. The first case is the Klein-Gordon field. In this case we get a double infinity of conserved quantities, since the theory is of second order in the derivatives of the field, and an algebra corresponding to the Heisenberg algebra in the continuum limit. The second  case is that of $N$ Fermi oscillators. In this case there are second class  constraints since the equations of motion are first order in the time derivatives. The result that we obtain here is analogous to the previous ones. In fact, with  the Fermi oscillators described by $N$ complex Grassmann variables, we get $2N$ conserved quantities in agreement with the order of the equations of motion ($d=1$) and the number of real coordinates ($2N$). Differently from the previous cases, the algebra of the generators is a Clifford one, due to the Grassmann character of the dynamical variables. However, in all  cases, the algebra of the conserved quantities is the  same as the algebra of the canonical variables describing the dynamical system. The third case is that of the Dirac equation, where we have both partial derivatives and anticommuting variables. We find again as many symmetries as the number of solutions of the field equations and the algebra of creation and annihilation Fermi operators.
 
 This paper is organised as follows: in Section 2 we prove our main result:   {\it any quadratic lagrangian depending on $N$ variables $q_a$ and their ordinary derivatives up to the order $m$, not giving rise to constraints in phase space, is invariant, up to total derivatives, under a translation of the dynamical variables $q_a$ by a solution of the equations of motion. These translations are generated by  $2mN$  constants of motion satisfying the Heisenberg algebra.} Since we are considering derivatives of an arbitrary order we make use of the Ostrogradsky formalism \cite{Ostrogradsky:1850os}. In particular we will prove that it is possible to choose the boundary conditions at $t=0$ for the solutions, in such a way that the algebra of the generators turns out to be universal, and independent of the details of the quadratic lagrangian.
 
 In Section 3 we make a similar analysis for the Klein-Gordon free field. This analysis is done in configuration space and we show the existence (as expected) of a double infinity of translational symmetries with related generators satisfying a continuum Heisenberg algebra.
 
 In Section 4 we study $N$ Fermi oscillators. This system is interesting because it is described by $N$ complex Grassmann coordinates and the resulting equations of motion are first order. As a consequence there are second class constraints. Also in this case we find that the number of conserved generators is given by the order of the equations of motion times the number of real degrees of freedom, that is $2N$.
 
 In Section 5 we deal with the Dirac equation, showing that  in this case too, as for the  Fermi oscillators, we have as many symmetries, corresponding to translations of the Dirac fields with solutions of the wave equation, as the number of independent solutions, since the Dirac equation is first order in the time derivatives.
 
 Finally we write our conclusions in Section 6.
 
We have also added two Appendices:
 in Appendix A Section 7 we   explicitly check  that the generators given in Section 2 are constant of motions. This calculation is interesting because it shows in an explicit way the role of the lagrangian as a quadratic function of the dynamical variables and their derivatives.
In Appendix  B Section 8  we present two particular cases of lagrangians giving rise to ordinary differential equations of fourth order in the derivatives. In both these examples we construct the solutions satisfying our particular boundary conditions at $t=0$. In the first case we also give  the explicit expressions of the generators.

\section{Quadratic lagrangians and translational symmetries}\label{sec:2}

To discuss the translational symmetries of a generic quadratic lagrangian  we will make use of their canonical description as done in \cite{Ostrogradsky:1850os} and reported in \cite{Whittaker:1937wh} (see also \cite{Barut:1962ba}). We will consider the following action depending on $N$ variables $q_a$ and on their time derivatives up to order $m$:
\be
S=\int dt \,L(q_a, q_a^{(1)} ,\cdots,q_a^{(m)}),\label{eq:2.1}
\ee
with
\be
q_a^{(i)}=\frac{d^iq_a}{dt^i} ,~~~i=1,\cdots m.
\ee
Varying the lagrangian we find:
\be
\delta L = \sum_{a=1}^N\left(\frac{\de L}{\de q_a}-\dot p_{a1}\right)\delta q_a+\frac d{dt}\sum_{a=1}^N\sum_{i=1}^m\left(p_{ai}\delta q_{ai}\right) ,\label{eq:6.35}
\ee
from which we get  the equations of motion:
\be
\frac{\de L}{\de q_a}-\dot p_{a1}=0,
\ee
where
\be
q_{ai}= q_a^{(i-1)}, ~~~i=1,\cdots, m,~~~ q_a^{(0)}=q_a \label{eq:2.5}
\ee
and
\bea
p_{a1}&= &\frac{\de L}{\de q_a^{(1)}}-\frac d{dt}\left(\frac{\de L}{\de q_a^{(2)}}\right)+\cdots\cdots\cdots+(-1)^{m-1}\frac {d^{m-1}}{dt^{m-1}}\left(\frac{\de L}{\de q_a^{(m)}}\right),\nn\\
p_{a2}&=& \frac{\de L}{\de q_a^{(2)}}-\frac d{dt}\left(\frac{\de L}{\de q_a^{(3)}}\right)+\cdots+(-1)^{m-2}\frac {d^{m-2}}{dt^{m-2}}\left(\frac{\de L}{\de q_a^{(m)}}\right),\nn\\
&&\cdots \cdots\cdots\cdots \cdots\cdots\cdots \cdots\cdots\nn,\\
p_{am}&=& \frac{\de L}{\de q_a^{(m)}}.\label{eq:2.6}
\eea
Notice that
\be
\dot q_{ai}=q_{a}^{(i)}.
\ee
Then, we define the hamiltonian of the system as
\be
H= \sum_{a=1}^N\sum_{i=1}^m p_{ai}\dot q_{ai} - L.
\ee
Its variation, using the eqs. \eq{2.6}, is given by
\be
\delta H =\sum_{a=1}^N\sum_{i=1}^m\left ( \delta p_{ai}\dot  q_{ai}-\dot p_{ai}\delta  q_{ai}\right)
\ee
and the Hamilton equations follow
\be
\frac{\de H}{\de p_{ai}}=\dot q_{ai},~~~\frac{\de H}{\de q_{ai}}=- \dot p_{ai}.
\ee
The Poisson brackets are defined in the usual way.

From now on, we will assume that the lagrangian \eq{2.1}
is quadratic   and that there is a term where the  highest derivative $q_a^{(m)}$, appears  as
\be
A_{ab} q_a^{(m)} q_b^{(m)},
\ee
giving rise to equations of motion  of degree $d=2m$. The reason for  this assumption is that we want to consider the case with no constraints in phase space. In fact, if
$q_a^{(m)}$ had not been coupled  with itself, but only to derivatives of lower degree, the momenta $p_{am}$ (see eq. \eq{2.6})
would  be  functions of the canonical coordinates $q_{ai}$ (see eq. \eq{2.5}), originating  constraints in phase space. We would like to stress that with  this assumption the equations of motion are of order $d=2m$  and, as such, they possess $dN$ independent solutions. Of  course, if there are no constraints, this can also be seen  from the previous hamiltonian analysis showing that our dynamical system is described by $2mN$  equations of first order, the Hamilton equations.

Since we are interested in  a quadratic lagrangian, the equations of motion are linear. Therefore the addition of a solution of the equations of motion to the variables $q_a$ leaves  the equations of  motion invariant. In this context we would like to prove the following statement:\\
 {\bf Any quadratic lagrangian depending on $N$ variables $q_a$ and their derivatives up to the order $m$, which does not give  rise to constraints in phase space, is invariant, up to total derivatives, under a translation of the dynamical variables $q_a$ by a solution of the equations of motion. These translations are generated by  $2mN$  constants of motion satisfying the Heisenberg algebra.}\\
 
 These symmetries are known for the  harmonic  (and the anharmonic) oscillator \cite{Lutzky:1978vi} and for its higher order analogue (Pais-Uhlenbeck oscillators), (see \cite{Andrzejewski:2014rza} and \cite{Masterov:2015ija}). They form a sub-algebra of the so called Newton-Hooke algebra \cite{Bacry:1968zf}.
 
Let us now prove the previous statement. Since the equations of motion have degree $d=2m$, there are $dN$ independent solutions that we will label as $f^{(pai)}, p =1,2$. For these solutions we will assume two kind of boundary conditions:  for the first $mN$ solutions
\be
f_{bj}^{(1ai)}(0)=\delta_{ij}\delta_{ab},~~~p_{bj}^{f^{(1ai)}}(0)=0,~~~i,j=1.\cdots, m,~~~a,b=1,\cdots,N\label{eq:2.12}
 \ee
 and for the other $mN$:
\be
f_{bj}^{(2ai)}(0)=0,~~~p_{bj}^{f^{(2ai)}}(0)=\delta_{ij}\delta_{ab}
~~~i,j=1.\cdots, m,~~~a,b=1,\cdots,N.\label{eq:2.13},
\ee
where, for all the solutions, we use \eq{2.5} to write the derivatives as:
\be
f_{ai} = f_a^{(i-1)}, ~~~i=1,\cdots, m,~~~ f_a^{(0)}=f_a
\ee
and  the $p_{ai}^{f^{(pbj)}}$ are defined by
\be
p_{ai}^{f^{(pbj)}}= p_{ai}|_{q_a=f_a^{(pbj)}}.
\ee
Notice that this choice of the boundary conditions means choosing the  functions $f_{bj}^{(1ai)}$ and the momenta $p_{bj}^{f^(2ai)}$ dimensionless. In particular this implies that the functions $f_{bj}^{(1ai)}$ and $f_{bj}^{(2ai)}$ have different dimensions as the corresponding momenta $p_{bj}^{f^(1ai)}$ and $p_{bj}^{f^(2ai)}$.

Then,  consider the following infinitesimal transformation
\be
\delta q_a= \sum_{b=1}^N\sum_{j=1}^m\sum_{p=1}^2\epsilon_{pbj} f_a^{(pbj)},\label{eq:2.15}
\ee
where $f_a$ are solutions of the equations of motion, Notice that  the parameters $\epsilon_{pbj}$ with $p=1$ and $p=2$ also have different dimensions. These translations are symmetries of the equations of motion due to their linearity. To prove that they are symmetries of the lagrangian (up to a total derivative),  we first  evaluate the variation of the generic term
\be
B_{ab}^{hk} q_a^{(h)} q_b^{(k)},
\ee
in two different ways: once integrating by parts the derivatives of $q_a$ and the other integrating by parts the derivatives of $\delta q_a$. We get:
\be
\delta(B_{ab}^{hk} q_a^{(h)} q_b^{(k)})=B_{ab}^{hk}\left[\frac d{dt}\left(\delta  q_a^{(h-1)} q_b^{(k)}+q_a^{(h)} \delta q_b^{(k-1)}\right)- \delta q_a^{(h-1)} q_b^{(k+1)}-q_a^{(h+1)} \delta q_b^{(k-1)}\right],
\ee
whereas integrating  by parts the derivatives of $q_a$ we get
\be
\delta(B_{ab}^{hk} q_a^{(h)} q_b^{(k)})=B_{ab}^{hk}\left[\frac d{dt}\left(\delta  q_a^{(h)} q_b^{(k-1)}+q_a^{(h-1)} \delta q_b^{(k)}\right)- \delta q_a^{(h+1)} q_b^{(k-1)}-q_a^{(h-1)} \delta q_b^{(k+1)}\right].
\ee
We see that the difference consists in the exchange $\delta q_a^{(h)}\leftrightarrow q_a^{(h)}$.
We can use these results and  \eq{6.35} to evaluate the variation of the lagrangian when integrating by parts the derivatives of $q_a$. Since the $f_a$'s are assumed to be solutions of the equations of motion, we get
\be
\delta L =\frac d{dt}\sum_{a,b=1}^N\sum_{i,j=1}^m\sum_{p=1}^2\epsilon_{pbj}\left(p_{ai}^{f^{(pbj)}} q_{ai}\right) .\label{eq:2.16}
\ee
 On the other hand, integrating by parts the derivatives of $\delta q_{ai}$ and assuming that the $q_a$'s satisfy the equations of motion, we get
 \be
\delta L =\frac d{dt}\sum_{a,b=1}^N\sum_{i,j=1}^m\sum_{p=1}^2\epsilon_{pbj}\left(p_{ai}\ f^{(pbj)}_{ai}\right). \label{eq:2.18}
\ee
Equating \eq{2.16} and \eq{2.18} it follows that the quantity:
\be
G^\delta=\sum_{a,b=1}^N\sum_{i,j=1}^m\sum_{p=1}^2\epsilon_{pbj}\left( p_{ai}^{f^{(pbj)}} q_{ai}-p_{ai}\ f^{(pbj)}_{ai}\right)
\ee
is a constant of motion. Therefore, we get the $2mN$ conserved generators:
\be
G^p_{bj}=\sum_{a=1}^N\sum_{i=1}^m\left( p_{ai}^{f^{(pbj)}} q_{ai}-p_{ai}\ f^{(pbj)}_{ai}\right).
\ee
 It is interesting to notice that the generators are nothing but the Wronskian of two solutions in the case of a second order differential equation. Therefore, the previous expression is a generalization of the Wronskian for higher order differential equations that can be obtained from a lagrangian.

 The Poisson brackets for two different solutions $f_a$ and $h_a$ are
\be
\{G_{ai}^p, G_{bj}^q\}=\sum_{c=1}^N\sum_{k=1}^m\left(- p_{ck}^{f^{(paj)}} f_{ck}^{(qbj)}+p_{ck}^{f^{(qbj)}} f^{(paj)}_{ck}\right).
\ee
Since the charges are constants of motion, we can evaluate their Poisson brackets at $t=0$.
Using the initial conditions \eq{2.12} and \eq{2.13}  we get
\be
\{G_{ai}^p, G_{bj}^q\}=\epsilon_{pq}\delta_{ab}\delta_{ij}.
\ee
It follows that the classical solutions $f_{ck}^{(qpj)}$ satisfy the condition
\be
\sum_{c=1}^N\sum_{k=1}^m\left(- p_{ck}^{f^{(1bj)}} f_{ck}^{(2ai)}+p_{ck}^{f^{(2ai)}} f^{(1bj)}_{ck}\right)=\delta_{ab}\delta_{ij}.
\ee

In an equivalent way, this result can   be obtained  noticing the relation between the generators and the canonical variables evaluated at $t=0$:
\be
G_{ai}^1 =-p_{ai}(0),~~~G_{ai}^2 = q_{ai}(0)
\ee
and using the Poisson brackets among the canonical coordinates.
A particular example is the free particle in $N$ dimensions. In this case $d=2$ and the two solutions are ($p^f=m\dot f$):
\be
f^{(1a)}_b=\delta^a_b,~~~f^{(2a)}_b= \delta^a_b\frac t m.
\ee
The corresponding conserved charges are the ordinary translations and the Galilei boosts:
\be
G_a^1=-p_a,~~~G_a^2= q_a-\frac {p_a} m t.
\ee
Another example is the one-dimensional harmonic oscillator. In this case:
\be
f^{(1)}=\cos\omega t,~~~f^{(2)}=
\frac 1{m\omega}\sin\omega t,
\ee
with generators:
\be
G^1= -m\omega q \sin\omega t- p\cos\omega t,~~~G^2= q\cos\omega t-\frac 1{m\omega} p\sin\omega t.
\ee

For the anharmonic oscillator, it is sufficient to substitute the trigonometric functions in the previous expressions with the hyperbolic ones. Two examples  with $d=4$  will be treated explicitly in Appendix B.

\section{Klein-Gordon equation}

We will now discuss the translational symmetries of a field theory described by a quadratic lagrangian. This time we will only consider  the specific case of the Klein-Gordon (KG) theory for a real scalar field. Of course we could consider the theory in a finite volume and describe the field in momentum space. This would lead to a series of decoupled harmonic oscillators that could be analysed according to the previous treatment and then taking the limit of infinite volume. However, it is more interesting to start directly in the continuum limit. The action describing this field is given by
\be
S=\int d^4 x\left(\frac 12 \de_\mu\phi\de^\mu\phi -\frac 12 m^2\phi^2\right)=\int d^4 x{\cal L}.
\ee
We will make use of the metric  $\eta_{\mu\nu}=(+1,-1,-1,-1)$.

For any fixed value of the spatial momentum $\vec k$, a generic solution of the KG  equation can be written as
\be
\frac 1{(2\pi)^{3/2}}e^{-i\vec k\cdot\vec x} \tilde f( x^0,\vec k), ~~~ \ddot {\tilde f} (x^0,\vec k)+\omega_{\vec k}^2 \tilde f( x^0,\vec k)=0,~~~\omega_{\vec k}=\sqrt{\vec k^{\,2}+m^2}.
\ee

In analogy with what we have done in eq. \eq{2.15}, we consider an arbitrary translation of the field $\phi(x)$ in terms of the solutions of the KG equation:
\be
\delta\phi(x)=\frac 1{(2\pi)^{3/2}}\sum_{p=1}^2\int d^3\vec k \, e^{-i\vec k\cdot\vec x} \tilde f_p( x^0,\vec k)\tilde\epsilon_p(\vec k),
\ee
where the index $p$ takes into account that the KG equation has two independent solutions according to the  boundary conditions, which will be chosen later on.
Using the properties of the  Fourier transform, we can write:
\be
\delta\phi(x)=\frac 1{(2\pi)^{3/2}}\sum_{p=1}^2\int d^3\vec y \,f_p(x^0,\vec x-\vec y) \,\epsilon_p(\vec y),
\ee
where $f_p(x)$ and $\epsilon_p(\vec x)$ are the Fourier transform of $\tilde f_p(x^0,\vec k)$ and $\tilde\epsilon_p(\vec k)$.  Since $f(x)$ satisfies the KG equation the same does $\delta\phi(x)$:
\be
(\de_\mu\de^\mu +m^2)\delta\phi(x)=0.
\ee
Using this result, the transformation $\delta\phi$ gives
\be
\delta S=\int d^4 x\,\de_\mu(\phi\de^\mu\delta\phi).
\ee
On the other hand, using the KG equation for $\phi$ we get the conserved quantity
\be
j_\mu=(\de_\mu\delta\phi)\phi- \delta\phi(\de_\mu\phi).
\ee
In a more explicit form
\be
j_\mu(x)= \sum_{p=1}^2 \int \frac{d^3\vec y}{(2\pi)^{3/2}} \left[\Pi_\mu^{f_p}(x^0,\vec x-\vec y) \phi(x)-f_p(x^0,\vec x-\vec y)\Pi_\mu(x)\right]\epsilon_p(\vec y),
\ee
where
\be
\Pi_\mu=\frac{\de{\cal L}}{\de(\de^\mu\phi)}=\de_\mu\phi,~~~\Pi_\mu^{f_p} =\de_\mu f_p.
\ee
Since $\epsilon_p(\vec y)$ is an arbitrary function,  we get an  infinite number of conserved currents
\be
\frac 1 {(2\pi)^{3/2}} \left[\Pi^{f_p}_\mu (x^0,\vec x-\vec y) \phi(x)-f_p(x^0,\vec x-\vec y)\Pi_\mu(x)\right].
\ee
Integrating over $d^3\vec x$,  we get a double  infinity of   conserved charges:
\be
G^p(\vec y) = \int  \frac{d^3\vec x}{(2\pi)^{3/2}}\left[\Pi_0^{f_p}(x^0,\vec x-\vec y) \phi(x)-f_p(x^0,\vec x-\vec y)\Pi_0(x)\right],~~~p=1,2,
\ee
where the two functions $f_p(x)$ are two  solutions corresponding to  independent boundary conditions.
The Poisson brackets among the conserved charges  are given by
\be
\{G^p(\vec y), G^q (\vec z)\}=\epsilon_{pq}\int  \frac{d^3\vec y}{(2\pi)^{3}}\left[f_1(x^0,\vec x-\vec y)\Pi_0^{f_2}(x^0,\vec x-\vec z)-
f_2(x^0,\vec x-\vec z) \Pi_0^{f_1}(x^0,\vec x-\vec y)\right].
\ee
Since the right hand side is time independent we can evaluate it at any time. Choosing the boundary conditions at $x^0=0$ as follows, 
\be
f_1(0,\vec x)=(2\pi)^{3/2}\delta^3(\vec x),~~~\Pi_0^{f_1}(0,\vec x)=0,
\ee
\be
f_2(0,\vec x)=0,~~~\Pi_0^{f_2}(0,\vec x)=(2\pi)^{3/2}\delta^3(\vec x),
\ee
we get:
\be
\{G^i(\vec y), G^j (\vec z)\}=\epsilon_{ij}\delta^3(\vec y-\vec z).\label{eq:3.27}
\ee
This can also be seen  using the relations between the charges and the canonical variables evaluated at $t=0$. We have:
\be
G^1(\vec x)=-\Pi_0(0, \vec x),~~G^2(\vec x)=\phi(0, \vec x)
\ee
Therefore, the Poisson brackets \eq{3.27} follow from the canonical Poisson brackets. 
The explicit expressions for the two solutions $f_p(x)$  are 
\be
f_1(x)=\frac 1{(2\pi)^{3/2}}\int d^3\vec k\, e^{i\vec k\cdot\vec x}\cos\omega_{\vec k} x^0,~~~f_2(x)=\frac 1{(2\pi)^{3/2}}\int d^3\vec k\, e^{i\vec k\cdot\vec x}\frac 1{\omega_{\vec k}}\sin\omega_{\vec k} x^0
\ee

\section{N Fermi oscillators}

 The case of $N$ Fermi oscillators is interesting for two reasons, the first  being that they are described by anti-commuting variables and the second because the lagrangian contains only first order derivatives, implying the presence of  second class constraints. Despite this fact, the number of conserved charges related to translations of the coordinates, performed with solutions of the equations of motion, is given again by the order of the equations of motion ($d=1$) times the number of real degrees of freedom ($2N$), that is $2N$, 

The $N$ Fermi oscillators can be described  by $N$ complex Grassmann coordinates satisfying the following lagrangian:
\be
L=\sum_{a=1}^N \left(\frac i2(\seta_a\dot\eta_a-\dot\eta_a^*\eta_a)-\omega_a\seta_a \eta_a\right).\label{eq:4.1}
\ee
We have
\be
p_a^\eta=\frac{\de L}{\de\dot\eta_a}=-\frac i2\seta_a,~~~\frac{\de L}{\de\eta_a}=\frac i 2\dot\eta_a^*+\omega_a\seta_a\label{eq:4.2},
\ee
\be
p_a^{\eta*}=\frac{\de L}{\de\dot\eta_a^*}=-\frac i2\eta_a,~~~\frac{\de L}{\de\seta_a}=\frac i 2\dot\eta_a-\omega_a\eta_a\label{eq:4.3},
\ee
giving the equations of motion
\be
\dot\eta^*_a=i\omega_a\seta_a,~~~\dot\eta_a=-i\omega_a\eta_a\label{eq:4.4}.
\ee

As in Section \ref{sec:2} we consider the transformations:
\be
\eta_a\to\eta_a+\delta\eta_a = \eta_a+\sum_{b=1}^N\epsilon_b f^{b}_a,~~~\seta_a\to \seta_a+\delta\seta_a=\seta_a+\sum_{b=1}^N\epsilon_b^* f^{b*}_a,
\ee
where the $\epsilon_a$ are anti-commuting parameters and the $f^a$   are $N$ bosonic independent solutions  of the equations of motion \eq{4.4}, satisfying the boundary conditions at $t=0$:
\be
f^a_b(0)=\delta_b^a\label{eq:4.6}.
\ee
The corresponding solutions
are
\be
f^a_b(t)=\delta_b^a e^{-i\omega_a t}\label{eq:4.7}.
\ee

Proceeding as in Section \ref{sec:2} we find that the quantity
\be
i\sum_{a,b=1}^N(\epsilon_a^* f_a^{b*}\eta_a-\seta_a \epsilon_a f_a^{b})
\ee
is a constant of motion. Since $\epsilon_a$ and $\epsilon_a^*$ are independent parameters, we get $2N$ conserved charges
\be
G^b=\sum_{a=1}^N f_a^{b*}\eta_a,~~~G^{b*}=\sum_{a=1}^N f_a^{b}\eta_a^*\label{eq:4.9}.
\ee
It follows  from eqs. \ref{eq:4.2} and \ref{eq:4.3} that we have $2N$ second class constraints:
\be
p_a^\eta=-\frac i2\seta_a,~~~p_a^{\eta*}=-\frac i2\eta_a.
\ee
The corresponding Dirac  brackets are
\be
\{\eta_a,\eta_b\}=\{\seta_a,\seta_b\}=0,~~~~\{\eta_a,\seta_b\}=-i\delta_{ab}\label{eq:4.11}.
\ee
Using the time independence of the generators $G^a$  and $G^{a*}$, and the boundary conditions on  $f_b^a$ we get
\be
\{G^a,G^{b*}\}=-i\delta_{ab},~~~\{G^a,G^{b}\}=\{G^{a*},G^{b*}\}=0.
\ee
After quantization these Poisson brackets go into  anticommutators with the correspondence rule: $[\cdot,\cdot]_+=i\{\cdot,\cdot\}$ \cite{Casalbuoni:1975bj}. Therefore, the quantum generators satisfy the algebra of the Fermi creation and annihilation operators.
Notice also that  the generators coincide with the canonical variables evaluated at $t=0$:
\be
G^b=\eta_b(0),~~~G^{b*}=\eta_b^*(0).
\ee

It is interesting to study the description of the Fermi oscillators in terms of real variables. Let us define:
\be
\xi_{1a}=\frac 1{\sqrt{2}} (\eta_a+\seta_a),~~~\xi_{2a}=\frac {-i}{\sqrt{2}} (\eta_a-\seta_a),
\ee
\be
f^{1a}_b=\frac 1{\sqrt{2}} (f^a_b+f^{a*}_b),~~~f^{2a}_b=\frac {-i}{\sqrt{2}} (f^a_b-f^{a*}_b).
\ee
In the $\xi_{ia}, i=1,2$ variables, the lagrangian \eq{4.1} becomes
\be
L= \frac i 2\sum_{i=1,2}\sum_{a=1}^N\xi_{ia}\dot\xi_{ia}-i\sum_{a=1}^N\omega_a
\xi_{1a}\xi_{2a}\label{eq:511}.
\ee
From the Dirac brackets \eq{4.11} for the complex variables, we get
\be
\{\xi_{ia},\xi_{jb}\}=-i\delta_{ij}\delta_{ab},~~~i,j=1,2,~~~a,b=1,\cdots,N,
\ee
showing that, after quantization,  the real  variables satisfy a Clifford algebra. 

From \eq{4.7} we get
\be
f^{1a}_b(t)=\sqrt{2}\delta_{ab}\cos\omega_at,~~~f^{2a}_b(t)=-\sqrt{2}\delta_{ab}\sin\omega_at,
\ee
with
\be
f^{1a}_b(0)=\sqrt{2}\delta_{ab},~~~f^{2a}_b(0)=0.
\ee
Then, using \eq{4.9}
\be
G^{1a}=\frac 1{\sqrt{2}}(G^a+G^{a*})= \frac 1{\sqrt{2}}\sum_{b=1}^N(f^{1a}_b\xi_{1b}+f^{2a}_b\xi_{2b}),
\ee
\be
G^{2a}=\frac {-i}{\sqrt{2}}(G^a-G^{a*})= \frac 1{\sqrt{2}}\sum_{b=1}^N(f^{1a}_b\xi_{2b}-f^{2a}_b\xi_{1b}),
\ee
we get
\be
\{G^{ia},G^{jb}\}=-i\delta_{ij}\delta_{ab}.
\ee
This can  also be seen by evaluating the generators at $t=0$. We get
\be
G^{1a}=\xi_{1a}(0),~~~G^{2a}=\xi_{2a}(0).
\ee
We see that the real generators satisfy the Clifford algebra after quantization.

\section{Dirac equation}

The final case we would like to consider is that of the Dirac equation, Let us consider the Dirac action for a fermi field with values in the Grassmann algebra (we will make use of the notations as in ref. \cite{Itzykson:1980}):
\be
S=\int~d^4x~\left(\frac i2(\bar\psi\gamma_\mu\de^\mu\psi -\de^\mu\bar\psi\gamma_\mu\psi) -m\bar\psi\psi\right),~~~
\hat v\equiv v_\mu\gamma^\mu.
\label{Dirac action}
\ee
The canonical momenta result to be
\be
\Pi_\psi=\frac{\de \LL}{\de\dot\psi}=-\frac i2\psi^\dagger,~~~~~~
\Pi_{\psi^\dagger}=\frac{\de\LL}{{\de\dot\psi}^\dagger}=-\frac i2\psi.\label{Dirac canonical momenta}
\ee
We see that there are second class constraints
\be
\chi=\Pi_\psi +\frac i2\psi^\dagger=0,~~~\chi^\dagger=\Pi_{\psi^\dagger}+\frac i2\psi=0.
\ee
The relevant Dirac brackets are
\be
\{\psi_\alpha(x^0,\vec x),\psi_\beta(x^0,\vec y\}=\{\psi^\dagger_\alpha(x^0,\vec x),\psi^\dagger_\beta(x^0,\vec y)\}=0,, 
\ee
\be
\{\psi_\alpha (x^0,\vec x),\psi^\dagger_\beta(x^0,\vec y)\}=-i\delta_{\alpha\beta}\delta^3(\vec x-\vec y).
\ee
Let us recall some properties of the spinors satisfying the Dirac equation in momentum space. Introducing the following notations:
\be
u^{(+)}(p,n) = u(p,n) ,~~~u^{(-)}(p,n) = v(p,n),
\ee
with $u(p,n)$ and $v(p,n)$ being the usual Dirac spinors, we have
\be
(\hat p -\epsilon m) u^{(\epsilon)}(p,n)=0,~~~\epsilon =(+,-),\label{eq:5.8}
\ee
\be
u^{(\epsilon')\dagger}( p_{\epsilon'},n') u^{(\epsilon)}( p_\epsilon,n)=\delta_{\epsilon'\epsilon}\delta_{n'n}\frac {E_p}m,
\ee
where
\be
p^\mu_\epsilon=(E_p,\epsilon\vec p),~~~E_{p}=\sqrt{|\vec p|^2+m^2}.
\ee

Let us now consider the  c-number  solution of the Dirac equation at fixed spatial momentum:
\be
 \psi^{(\epsilon, n)}(x,\vec p)=\sqrt{\frac m{E_p}}e^{-i\epsilon E_p x^0 +i\vec p\cdot\vec x} u^{(\epsilon)}(\epsilon p,n).
 \ee
  
A variation of the lagrangian corresponding to a translation of the Dirac field with a generic combination of the solutions $ \psi^{(\epsilon, n)}(x,\vec p)$ is given by
  \be
  \delta\psi(x)= \sum_{\epsilon,n}\int \frac {d^3\vec p}{\sqrt{(2\pi)^3}} \psi^{(\epsilon, n)}(x,\vec p)\epsilon_{\epsilon,n}(\vec p), \ee
 where $\epsilon_{\epsilon,n}(\vec p)$ are arbitrary complex  parameters with values in the Grassmann algebra. Assuming that also the field $\psi(x)$ satisfies the Dirac equation we find the conserved current
 \be
 \de^\mu\left(\delta\bar\psi\gamma _\mu\psi -\bar\psi\gamma _\mu\delta\psi\right) =0.
 \ee
 Since the Grassmann parameters are arbitrary, we find two conserved currents
 \be
 \frac {1}{\sqrt{(2\pi)^3}} \bar\psi^{(\epsilon, n)}(x,\vec p) \gamma_\mu\psi(x),~~~
 \frac {1}{\sqrt{(2\pi)^3}} \bar\psi(x) \gamma_\mu\psi^{(\epsilon, n)}(x,\vec p). 
 \ee
 Therefore, we get the conserved generators
 \be
 G^{(\epsilon, n)}(\vec p) = \int \frac {d^3\vec x}{\sqrt{2\pi^3}} \psi^{(\epsilon, n)\dagger}(x,\vec p) \psi(x),~~~
 G^{(\epsilon, n)\dagger}(\vec p) = \int \frac {d^3\vec x}{\sqrt{2\pi^3}} \psi(x)^\dagger \psi^{(\epsilon, n)}(x,\vec p) . \ee
 
 The Dirac brackets of the generators are 
 \be
 \{ G^{(\epsilon', n')}(\vec p') , G^{(\epsilon, n)}(\vec p) \}= \{ G^{(\epsilon', n')\dagger}(\vec p') , G^{(\epsilon, n)\dagger}(\vec p) \} =0,
 \ee
 \be
 \{ G^{(\epsilon', n')}(\vec p') , G^{(\epsilon, n)\dagger}(\vec p) \} =-i\delta_{\epsilon'\epsilon}\delta_{n'n}\delta^3(\vec p'-\vec p).
 \ee
After quantization these are the anticommutation relations for the Dirac creation and annihilation operators.

One could also consider the generators in coordinate space:
\be
 G^{\epsilon', n'}(\vec x)= \int \frac {d^3{\vec p}}{\sqrt{2\pi^3}}e^{-i\vec p\cdot\vec x}G^{\epsilon, n}(\vec p).
 \ee
 In this case the Poisson brackets are given by
 \be
 \{ G^{\epsilon', n'}(\vec x') , G^{\epsilon, n\dagger}(\vec x) \} =-i\delta_{\epsilon'\epsilon}\delta_{n'n}\delta^3(\vec x'-\vec x).
 \ee
We see that  the number of conserved generators is equal to that of independent solutions of the Dirac equation, since the Dirac equation is first order in the time derivatives,

\section{Conclusions}

In this paper we have proved that a quadratic lagrangian,  with no constraints, containing  ordinary time derivatives up to the order $m$ of $N$ dynamical variables has $2mN$ symmetries consisting in the translation of the variables with solutions of the equations of motion.   The generators of these transformations have then been constructed explicitly and we have shown that, using convenient conditions at $t=0$, the resulting algebra of the generators is the Heisenberg algebra. We conjecture that the result is more   generally valid with respect to what we have proven explicitly. To this end we have considered the case of the Klein-Gordon lagrangian, which gives rise to a partial differential equation of the second order in the  derivatives. In this case we find, as expected a double infinity of translational symmetries. We have also considered the case of $N$ Fermi oscillators. This case is interesting because it is described by Grassmann variables and because the lagrangian has second class constraints.  Here too we find that the translations of the dynamical variables are symmetries and that their number is given by the degree of the equations of motion times the number of real degrees of freedom. Analogous results have been found for the Dirac lagrangian.

In the case of the KG lagrangian, the previous result can be expected simply from the observation that, the KG field can after all be seen as a collection of infinite harmonic oscillators which satisfy our theorem. The same happens for the Dirac case, where the field can be considered as an infinite collection of  Fermi oscillators.

The symmetries discussed in this paper are in one to one correspondence with the solutions of the equations of motion of any given model  and, therefore, they  cannot give any restriction on the space of  solutions. 

A general  discussion of the physical relevance of the symmetries presented here it is not easy since it depends a lot on their interpretation which, in turn, depends on the physical system one is considering. To be more explicit, let us think to a system described by space variables. In this case the translations discussed here  are related to the structure of the space itself. For instance, in the case of a free non relativistic particle, our analysis leads to space translations and boosts invariances, which are typical of a flat non relativistic space-time. Analogously, if we consider a harmonic or anharmonic oscillator, our  analysis leads to a subgroup of the Newton-Hooke group (precisely the translations). It turns out that these are symmetries of  contractions of the (A)dS group and, in particular the translations, made with the solutions of the equations of motion of the oscillator, represent movements along the curved directions \cite{Bacry:1968zf}. 

On the other hand, if we consider a field theory, our symmetries are internal symmetries.
They do not touch the space time structure but add   the field variables functions that are non-linear in space time.  For instance, this is relevant for the massless KG equation which is invariant (at the classical level) under a constant translation. Also the polynomial shift symmetries considered in
\cite{Griffin:2014bta} can be covered by our analysis.

Finally, let us emphasise that the quadratic lagrangian  possess  as many non trivial  constants of motion (the generators of the translational symmetries) which can be very useful in discussing the physical properties of a dynamical system.

\acknowledgments{
We would like to thank Joaquim Gomis for the many discussions and for his very  useful comments about this paper.
}

\section{Appendix A:  Check of the time independence of the charges $G^f$}

It is interesting to  explicitly check the time independence of the charges $G^f$ defined in Section \ref{sec:2}, because this shows how this result is directly related to the  lagrangian and  the hamiltonian being  quadratic functions of the dynamical variables. 

Assume that $f$ is a solution of the equations of motion, we want to show that
 the quantity
\be
G^f=\sum_{a=1}^N\sum_{i=1}^m\left( p_{ai}^{f} q_{ai}-p_{ai}\ f_{ai}\right),
\ee
is a constant of motion. We have,
\be
    \dot G^f=\sum_{a=1}^N\sum_{i=1}^m\left( \dot p_{ai}^f q_{ai}+ p_{ai}^f\dot  q_{ai}-\dot  p_{ai}f_{ai}- p_{ai}\dot  f_{ai}\right).
\ee
Using the Hamilton equations we can write this expression in the following form
\be
  \dot G^f=\sum_{a=1}^N\sum_{i=1}^m\left(-\frac {\de H^f}{\de f_{ai}} q_{ai}+ p_{ai}^f\frac{\de H}{\de p_{ai}}+\frac{\de H}{\de q_{ai}} f_{ai}- p_{ai}\frac{\de H^f}{\de p_{ai}^f}\right),
\ee
where
\be
H^f=H(f_{ai},p^f_{ai}).
\ee
Introducing the following quantities:
\be
X_{ai}=(q_{ai},p_{ai}),~~~X_{ai}^f=(f_{ai},p_{ai}^f),
\ee
we get
\be
  \dot G^f=-\sum_{a=1}^N\sum_{i=1}^m\left( \frac {\de H^f}{\de X_{ai}^f}X_{ai}-\frac{\de H}{\de X_{ai}}\ X^f_{ai})\right).
\ee
The hamiltonian is a quadratic function of the variables $X_{ai}$, since the lagrangian is quadratic in its variables. Therefore we have
\be
H=\frac 12\sum_{abij} X_{ai}C_{aibj}X_{bj}.
\ee
Using this expression for the hamiltonian and the symmetry of $C_{aibj}$ in the exchange $ai \leftrightarrow bj$, it is easy to verify that $\dot G^f=0$.

\section{Appendix B:  Two simple examples}

We will  now  consider two examples of lagrangians with higher order derivatives to illustrate our procedure. The first and simplest example  is the following:
\be
L=\frac a2 \ddot q^2,
\ee
with equation of motion $d^4 q/dt^4=0$. The momenta are:
\be
p_1=-a \dddot q,  ~~~  p_2=a\ddot q.
\ee
The four independent solutions with the appropriate boundary conditions, given in Table 1, are:
\be
f^{(11)}_1 =1,~~~f^{(12)}_1 =t,~~~f^{(21)}_1 =-\frac 1{6a} t^3,~~~f^{(22)}_1 =\frac 1{2a} t^2,
\ee
\be
f^{(11)}_2 =0,~~~f^{(12)}_2 =1,~~~f^{(21)}_2 =-\frac 1{2a} t^2,~~~f^{(22)}_2 =\frac 1{a} t
\ee
and
\be 
p^{f^{(11)}_1}_1=0,~~~p^{f^{(12)}_1}_1=0,~~~p^{f^{(21)}_1}_1=1,~~~p^{f^{(22)}_1}_1=0,
\ee
\be 
p^{f^{(11)}_1}_2=0,~~~p^{f^{(12)}_1}_2=0,~~~p^{f^{(21)}_1}_2=-t,~~~p^{f^{(22)}_1}_2=1.
\ee

The generators are given by
\be
G_1^1=-p_1,~~~G_2^1=-p_1t-p_2,\ee
\be
G_1^2=q_1-q_2 t +\frac 1{6a} p_1 t^3+\frac 1{2a} p_2 t^2,~~~G_2^2=q_2-\frac 1{2a}p_1 t^2-\frac 1 a p_2 t.
\ee
The hamiltonian is
\be
H = p_1q_2+\frac 1{2a} p_2^2,
\ee
giving rise to the following hamilton equations 
\be
\dot q_1=q_2,~~~\dot q_2=\frac 1 a p_2,~~~\dot p_1=0,~~~\dot p_2= -p_1.
\ee
It   is easy to check that all the generators and the hamiltonian are conserved quantities
\be
\dot H=\dot G_i^j=0.
\ee
Finally the Poisson brackets of the generators with the hamiltonian are:
\be
\{G_1^1,H\}=0,~~~\{G_2^1,H\}=-G_1^1,~~~\{G_1^2,H\}=G_2^2,~~~\{G_2^2,H\}=-\frac 1 a G_2^1.
\ee

The second example that we will consider is the Pais-Uhlenbeck oscillator  (see \cite{Andrzejewski:2014rza} and \cite{Masterov:2015ija}): 
\be
L=\frac 12 a\ddot q^2+\frac 12 b \dot q^2+\frac 12 c q^2=a\left(\frac 12 \ddot q^2-\frac 12 (\omega_1^2+\omega_2^2) \dot q^2+\frac 12\omega_1^2\omega_2^2  q^2\right),
\ee
with
\be
b= -a(\omega_1^2+\omega_2^2),~~~c=a\omega_1^2\omega_2^2,
\ee
where we have  assumed $a,c>0$, $b<0$ and $b^2>4ac$. Furthermore we have introduced the frequencies:
\be
\omega_1^2=\frac{-b+\sqrt{b^2-4ac}}{2a}, ~~~\omega_2^2=\frac{-b-\sqrt{b^2-4ac}}{2a}.
\ee
The canonical variables are given by
\be
q_1=q,~~~q_2=\dot q,~~~p_1= b\dot q-a \dddot q,  ~~~  p_2=a\ddot q.
\ee
\begin{table}[h]
\caption{In this Table we give the values of the components at $t=0$ of the functions $f^{(p,i)}$ and their derivatives .}
\begin{center}
\begin{tabular}{|c||c|c|c|c|}
\hline
&&&&\\
 & $f^{(p,i)}_1=f^{(p,i)}$ & $f^{(p,i)}_2=\dot f^{(p,i)}$ & $p_1^f=b\dot f^{(p,i)}-a \dddot f^{(p,i)}$ &  $p_2^f=a\ddot f^{(p,i)}$ \\
&&&&\\
 \hline\hline
 $f^{(11)}$& 1 & 0 & 0& 0\\
 \hline
 $f^{(12)}$& 0 & 1 & 0& 0\\
 \hline
 $f^{(21)}$& 0 & 0 & 1& 0\\
 \hline
$f^{(22)}$& 0 & 0 & 0& 1\\
 \hline
\end{tabular}
\end{center}
\label{default}
\end{table}

The equations of motion factorise as follows:
\be
a\left(\frac {d^2}{dt^2}+\omega_1^2\right)\left(\frac {d^2}{dt^2}+\omega_2^2\right)q=0.
\ee
The general solution of this equation is:
\be
f= a_1 c_1+a_2 c_2+b_1 s_1+b_2 s_2,
\ee
where
\be
c_i=\cos\omega_i t,~~~s_i=\sin\omega_i t,~~~i=1,2.
\ee
Given the boundary conditions in Table I
 the four independent solutions of the equations of motion  are:
$$
f^{(11)}= -\frac{\omega_2^2}{\omega_1^2-\omega_2^2} c_1 +\frac{\omega_1^2}{\omega_1^2-\omega_2^2} c_2,
$$
$$
 f^{(12)}= \frac{\omega_1}{(\omega_1^2-\omega_2^2)} s_1 -\frac{\omega_2}{(\omega_1^2-\omega_2^2)} s_2,
$$
$$
f^{(21)}= \frac{1}{a\omega_1(\omega_1^2-\omega_2^2)} s_1 -\frac{1}{a\omega_2(\omega_1^2-\omega_2^2)} s_2,
$$
$$
 f^{(22)}= -\frac{1}{a(\omega_1^2-\omega_2^2)} c_1 +\frac{1}{a(\omega_1^2-\omega_2^2)} c_2.
$$

\end{document}